\documentclass{iopart}
\usepackage{graphicx}  
\usepackage{latexsym}
\usepackage{amssymb}

\jl{6}        
\eqnobysec    

\def\beq{\begin{equation}}
\def\eeq{\end{equation}}
\def\rmd{{\rm d}} 

\begin{document}

\title
[Emitting source in Schwarzschild spacetime radiation field]
{The signal from an emitting source moving in a Schwarzschild 
spacetime under the influence of a radiation field}

\author{
Donato Bini${}^*{}^\S$,
Maurizio Falanga${}^\parallel$,
Andrea Geralico${}^\ddag{}^\S$ and 
Luigi Stella$^\dag$
}
\address{
  ${}^*$\
Istituto per le Applicazioni del Calcolo ``M. Picone,'' CNR, I-00161 Rome, Italy
}
\address{
  ${}^\S$\
  ICRA, University of Rome ``La Sapienza,'' I-00185 Rome, Italy
}
\address{
  ${}^\parallel$\
  International Space Science Institute (ISSI), CH-3012 Bern, Switzerland
}
\address{
  ${}^\ddag$\
  Physics Department, University of Rome ``La Sapienza,'' I-00185 Rome, Italy
}
\address{
  ${}^\dag$\
   Osservatorio Astronomico di Roma, via Frascati 33, I-00040 Monteporzio Catone (Roma), Italy
}

\begin{abstract}
The motion of matter immersed in a radiation field is affected by radiation drag, as a result 
of scattering or absorption and re-emission. The resulting friction-like drag, also known
as Poynting-Robertson effect, has been recently studied in the general relativistic background 
of the Schwarzschild and Kerr metric, under the assumption that all photons in the radiation 
field possess the same angular momentum. We calculate here the signal produced 
by an emitting point-like specific source moving in a Schwarzschild spacetime under the influence 
of such a radiation field. We derive the flux, redshift factor and solid angle of the 
hot spot as a function of (coordinate) time, as well as the time-integrated image of the 
hot spot as seen by an observer at infinity.
The results are then compared with those for a spot moving on a circular geodesic in a
Schwarzschild metric. 

\end{abstract}

\pacno{04.20.Cv}

\section{Introduction}

In previous works \cite{bijanste,pr2} we studied the motion of test particles 
in the gravitational background of a black hole under the influence of a superimposed
radiation field. The particle-radiation interaction is assumed to take place through 
Thomson scattering only. Radiation exerts a drag force on the particle's motion, a well-known
effect that in the Newtonian regime is often referred to as the Poynting-Robertson 
effect \cite{poynting,robertson}. We limited ourselves to the case of particles moving in 
the equatorial plane of Schwarzschild and Kerr backgrounds, immersed in an (equatorial) 
radiation field composed of photons having the same specific angular momentum and traveling 
along geodesics. In Ref. \cite{bijanste} the simplest case 
was considered of a radiation field with zero angular momentum, i.e. photons moving 
in a purely radial direction with respect to the locally nonrotating frames, naturally 
associated with the family of zero angular momentum observers (ZAMOs). 
This analysis was extended in Ref. \cite{pr2} to the case of a radiation field characterized
by photons possessing all the same, non-null specific angular momentum.
We found that those particles which do not escape to infinity are attracted to a
single critical radius outside the horizon where they stay at rest with respect to ZAMOs, if the radiation
field and the black hole have zero angular momentum, or move in a circular orbit, 
if a non-null angular momentum characterizes the radiation field and/or 
the black hole. 

In the present paper we consider an emitting source (hot spot) which 
moves in a Schwarzschild metric under the influence of a radiation field, and calculate
the correponding signal seen by an observer at infinity. In particular we derive the flux, 
redshift factor and solid angle of the hot spot as a function of (coordinate) time, as well as the 
time-integrated image of the hot spot in the observer's sky.
The results are then compared with those for a spot moving on a circular geodesic in a
Schwarzschild field. This analysis will then be extended to the more complex case of a 
Kerr field in a forthcoming paper.

\section{Motion in the Schwarzschild spacetime}

Consider a Schwarzschild spacetime, whose line element written in standard coordinates is given by
\beq 
\label{metric}
\rmd  s^2 = -N^2\rmd t^2 + N^{-2} \rmd r^2 
+ r^2 (\rmd \theta^2 +\sin^2 \theta \rmd \phi^2)\,,
\eeq
where $N=(1-2M/r)^{1/2}$ denotes the lapse function, and introduce the usual orthonormal frame adapted to the Zero Angular Momentum Observers (ZAMOs) following the time lines
\beq
\label{frame}
n\equiv 
e_{\hat t}=N^{-1}\partial_t\,, \quad
e_{\hat r}=N\partial_r\,, \quad
e_{\hat \theta}=\frac{1}{r}\partial_\theta\,, \quad
e_{\hat \phi}=\frac{1}{r\sin \theta}\partial_\phi\,,
\eeq
where $\{\partial_t, \partial_r, \partial_\theta, \partial_\phi\}$ is the coordinate frame.

\subsection{Circular geodesic motion}

Circular geodesic motion of a test particle in the equatorial plane $\theta=\pi/2$ at $r=r_0$ is characterized by the 4-velocity
\beq
\label{Ugeo}
U=U_K=\gamma_K (n \pm \nu_K e_{\hat \phi})\,,
\eeq
where the Keplerian value of speed ($\nu_K$), the associated Lorentz factor ($\gamma_K$) and angular velocity ($\zeta_K$) are given by
\beq\fl\qquad
\label{nuKdef}
\nu_K=\sqrt{\frac{M}{r_0-2M}}\,,\qquad 
\gamma_K=\sqrt{\frac{r_0-2M}{r_0-3M}}\,,\qquad 
\zeta_K=\sqrt{\frac{M}{r_0^3}}\,.
\eeq
The $\pm$ signs in Eq.~(\ref{Ugeo}) correspond to co-rotating $(+)$ or counter-rotating $(-)$ orbits with respect to increasing values of the azimuthal coordinate $\phi$ (counter-clockwise motion as seen from above). 

The parametric equations of $U_K$ are
\beq
\label{circgeos}
t_K=t_0+\Gamma_K \tau\,,\quad
r_K=r_0\,,\quad 
\theta_K=\frac{\pi}{2}\,,\quad
\phi_K=\phi_0\pm \Omega_K \tau\,,
\eeq
where 
\beq
\Gamma_K=\sqrt{\frac{r_0}{r_0-3M}}\,, \qquad
\Omega_K=\frac{1}{r_0}\sqrt{\frac{M}{r_0-3M}}\,.
\eeq

\subsection{Test particle undergoing Poynting-Robertson effect}

Consider now a test particle in arbitrary motion on the equatorial plane $\theta=\pi/2$, i.e., with 4-velocity and 3-velocity with respect to the ZAMOs respectively
\beq\label{polarnu}
\fl\quad 
U=\gamma(U,n) [n+ \nu(U,n)]\,,\quad
\nu(U,n)\equiv \nu^{\hat r}e_{\hat r}+\nu^{\hat \phi}e_{\hat \phi} 
=  \nu \sin \alpha\, e_{\hat r}+ \nu \cos \alpha\, e_{\hat \phi}  \,,
\eeq
where $\gamma(U,n)=1/\sqrt{1-||\nu(U,n)||^2}$ is the Lorentz factor and the abbreviated notation $\nu^{\hat a}=\nu(U,n)^{\hat a}$ has been used. In a similarly abbreviated notation, $\nu = ||\nu(U,n)||\ge0$ and $\alpha$ are the magnitude of the spatial velocity $\nu(U,n)$ and its
polar angle  measured clockwise from the positive $\phi$ direction in the $r$-$\phi$ tangent plane.
Note that $\sin\alpha=0$ (i.e., $\alpha=0,\pi$) corresponds to purely azimuthal motion of the particle with respect to the ZAMOs, while $\cos\alpha=0$
(i.e., $\alpha=\pm \pi/2$) corresponds to (outward/inward) purely radial motion with respect to the ZAMOs.

Let the particle be accelerated by a test radiation field propagating in a general direction on the equatorial plane. 
The corresponding equations for $\nu$ and $\alpha$ as derived in \cite{pr2} are given by
\begin{eqnarray}
\label{dnuandalphadtau}
\frac{\rmd \nu}{\rmd \tau}
&=& 
-\frac{\sin\alpha}{\gamma}\frac{M}{r^2N}
   \nonumber \\
&& + \frac{A}{N^2r^2|\sin \beta|} 
[\cos(\alpha-\beta) -\nu][1-\nu\cos(\alpha-\beta)] \,, \nonumber\\
\frac{\rmd \alpha}{\rmd \tau}
&=& \frac{\gamma\cos \alpha}{\nu}\frac{N}{r}\left(\nu^2-\frac{M}{rN^2}\right) 
\nonumber \\
&&
+\frac{A}{\nu }  
\frac{[1-\nu\cos(\alpha-\beta)] }{N^2r^2|\sin \beta|}  \sin(\beta -\alpha)\,,
\end{eqnarray}
where $\tau$ is the proper time parameter along $U$, $A$ is a positive constant for a given fixed radiation field and  
\beq
\label{cosbeta}
\cos \beta 
=\frac{b_{\rm(rad)}N}{r}\,,
\eeq
the constant $b_{\rm(rad)}$ being the photon impact parameter defined as the ratio between the conserved angular momentum and the energy associated with the rotational and timelike Killing vector fields, respectively.
In the zero angular momentum limit $b_{\rm(rad)}\to 0$, $\cos\beta\to0$, then 
$\sin\beta\to\pm1$,
$\cos(\alpha-\beta)\to \pm\sin\alpha$,
$\sin(\beta-\alpha)\to \pm\cos\alpha$ and
the two equations (\ref{dnuandalphadtau}) reduce to those in \cite{bijanste}.

Finally, we have in addition the remaining equations
\beq
\frac{\rmd t}{\rmd \tau}=
\frac{\gamma}{N}\,,
\qquad
\frac{\rmd r}{\rmd \tau}=
\gamma N\nu\sin\alpha\,,
\qquad
\frac{\rmd \phi}{\rmd \tau}=
\frac{\gamma}{r}\nu\cos\alpha
\,.
\eeq

The system of four differential equations for $\nu$, $\alpha$, $r$ and $\phi$ admits
a critical solution at a radial equilibrium which corresponds to a circular orbit of constant radius $r=r_0$, constant speed $\nu=\nu_0$, and constant angles $\beta=\beta_0$, $\alpha=\alpha_0$.
The constancy of the radius requires  $\sin\alpha_0=0$, $\cos\alpha_0=\pm1$  and therefore
$\sin(\beta_0-\alpha_0)=\cos\alpha_0 \sin\beta_0$ and
$\cos(\alpha_0-\beta_0)=\cos\alpha_0 \cos\beta_0$.
The force balance equation for the critical circular orbits is then 
\begin{eqnarray}
\label{forcebalance2s}
 N\gamma_0^{3}\left( 1-\frac{\nu_0^2}{\nu_K^2}\right)
&=&
 {\rm sgn}(\sin \beta_0)\frac{A}{ M }  
\,, 
\end{eqnarray}
with
\beq
\label{nu0limit}
\pm\nu_0 = \cos \beta_0 
= \frac{ b_{\rm(rad)} N}{r_0}\quad
\rightarrow 
\quad
  \gamma_0= 1/|\sin\beta_0|
\,.
\eeq
In the case $b_{\rm(rad)}=0$ (i.e. $\nu_0=0$, $\gamma_0=1$) and $\sin\beta_0>0$ of purely radial outward photon motion, the previous equation reduces to the result of Ref. \cite{bijanste}
\beq
\label{rcrit_zeroang}
\frac{A}{M} = N = \left(1-\frac{2M}{r_0}\right)^{1/2}\,,
\eeq 
which requires  $A/M<1$ for a solution to exist.

\subsection{Radiation from the spot}

Let the spot to radiate isotropically in its own rest frame.
Consider then a (geodesic) photon connecting the emitter world line with the observer world line, i.e.,
\beq
K=\Gamma_{\rm (ph)}[\partial_t +\zeta^a_{\rm (ph)} \partial_a]\,.
\eeq
For a general motion we have 
\begin{eqnarray}
\label{Kcompts}
K^t=\frac{\rmd t}{\rmd \lambda}&=& 
\frac{E}{N^2}
=\Gamma_{\rm (ph)}\,,\nonumber \\
K^r=\frac{\rmd r}{\rmd \lambda}&=&
\epsilon_r \frac{E}{r^2}\sqrt{R}
=\Gamma_{\rm (ph)}\zeta^r_{\rm (ph)}\,,\nonumber \\
K^\theta=\frac{\rmd \theta}{\rmd \lambda}&=&
\epsilon_\theta \frac{E}{r^2}\sqrt{\Theta}
=\Gamma_{\rm (ph)}\zeta^\theta_{\rm (ph)}\,,\nonumber \\
K^\phi=\frac{\rmd \phi}{\rmd \lambda}&=& 
\frac{E}{r^2}\frac{b}{\sin^2\theta}
=\Gamma_{\rm (ph)}\zeta^\phi_{\rm (ph)}\,,
\end{eqnarray}
where $\lambda$ is an affine parameter, $\epsilon_r$ and $\epsilon_\theta$ are sign indicators, 
\begin{eqnarray}
R&=& r^4-r^2N^2(b^2+q^2)=r[r^3-(r-2M)(b^2+q^2)]\,,\nonumber\\
\Theta&=& q^2-b^2\cot^2\theta\,,
\end{eqnarray}
and the following notation has been introduced 
\beq
b=\frac{L}{E}\,,\qquad 
q^2=\frac{{\mathcal K}}{E^2}\,,
\eeq
$E$, $L$ and ${\mathcal K}$ being conserved Killing quantities.
Note that $q^2$ can take any value. 

The geodesic equations can be formally integrated by eliminating the affine parameter as follows
\begin{eqnarray}
\label{nullgeos}
&&\epsilon_r\int^{r}\frac{\rmd r}{\sqrt{R(r)}}=
\epsilon_\theta\int^{\theta}\frac{\rmd \theta}{\sqrt{\Theta(\theta)}}\ , \nonumber\\
&&t=\epsilon_r\int^{r}\frac{r^2}{N^2\sqrt{R(r)}}\,\rmd r\ , \nonumber\\
&&\phi=b\epsilon_\theta\int^{\theta}\frac{\rmd\theta}{\sin^2\theta\sqrt{\Theta(\theta)}}\ .
\end{eqnarray}
The integrals are along the path of motion.

\section{Energy shift}

We consider here the case of an emitter which moves on the equatorial plane of a Schwarzschild spacetime, the observer  at rest (very far from the origin) at a point not necessarily belonging to the equatorial plane, the photon connecting the two world lines, namely
\begin{eqnarray}
U_{\rm (em)}&=&\Gamma_{\rm (em)}[\partial_t +\zeta^r_{\rm (em)} \partial_r+\zeta^\phi_{\rm (em)} \partial_\phi]\,, \nonumber \\
U_{\rm (obs)}&=&\partial_t\,, \nonumber \\
K&=&\Gamma_{\rm (ph)}[\partial_t +\zeta^r_{\rm (ph)} \partial_r +\zeta^\theta_{\rm (ph)} \partial_\theta+\zeta^\phi_{\rm (ph)} \partial_\phi]\ .
\end{eqnarray}

The energy of the photon at the emission point $P_{\rm (em)}$, as measured by $U_{\rm (em)}$, is
\begin{eqnarray}
\fl\qquad 
E_{\rm (em)}& \equiv & E(K,U_{\rm (em)}) =-K\cdot U_{\rm (em)}|_{P_{\rm (em)}}\nonumber \\
\fl\qquad 
&=&- \Gamma_{\rm (ph)} \Gamma_{\rm (em)} [-N^2+N^{-2}\zeta^r_{\rm (ph)}\zeta^r_{\rm (em)}+r^2\zeta^\phi_{\rm (ph)}\zeta^\phi_{\rm (em)}]|_{P_{\rm (em)}}\,,
\end{eqnarray}
while the one observed at the point $P_{\rm (obs)}$ by $U_{\rm (obs)}$ is
\beq
E_{\rm (obs)}\equiv E(K,U_{\rm (obs)}) =-K\cdot U_{\rm (obs)}|_{P_{\rm (obs)}}=-E|_{P_{\rm (obs)}}\, .
\eeq

Therefore, the ratio $E_{\rm (obs)}/E_{\rm (em)}$ is
\begin{eqnarray}
\label{redshift}
g\equiv
\frac{E_{\rm (obs)}}{E_{\rm (em)}}&=&
\frac{\sqrt{N^2-N^{-2}\zeta^r_{\rm (em)}{}^2-r^2\zeta^\phi_{\rm (em)}{}^2}}
{1-N^{-4}\zeta^r_{\rm (ph)}\zeta^r_{\rm (em)}-b \zeta^\phi_{\rm (em)}}
\end{eqnarray}
with all quantities evaluated ad the emission point.
Usually one also introduces  the \lq\lq redshift" parameter $z$
\beq
z=\frac{E_{\rm (em)}-E_{\rm (obs)}}{E_{\rm (obs)}}\,,\qquad 
g=(z+1)^{-1}\ .
\eeq
If the emitter is in a geodesic circular orbit, we have $\zeta^r_{\rm (em)}=0$ and $\zeta^\phi_{\rm (em)}=\pm\zeta_K$, so that 
\beq
g=\frac{\sqrt{1-\displaystyle \frac{3M}{r_{\rm (em)}}}}
{1\mp b \sqrt{\displaystyle \frac{M}{r^3_{\rm (em)}}}}\ .
\eeq

\section{Ray-tracing}

Consider a photon emitted at the point $r_{\rm em}$, $\theta_{\rm em}$ and $\phi_{\rm em}$ at the coordinate time $t_{\rm em}$, which reaches an observer located at $r_{\rm obs}$, $\theta_{\rm obs}$ and $\phi_{\rm obs}$ at the coordinate time $t_{\rm obs}$.
The photon trajectories originating at the emitter must satisfy the following integral equation
\beq
\label{eqrth}
\epsilon_r\int_{r_{\rm em}}^{r}\frac{\rmd r}{\sqrt{R(r)}}=
\epsilon_\theta\int_{\theta_{\rm em}}^{\theta}\frac{\rmd \theta}{\sqrt{\Theta(\theta)}}\ ,
\eeq 
as from Eq. (\ref{nullgeos}).
The signs $\epsilon_r$ and $\epsilon_\theta$ change when a turning point is reached.
Turning points in $r$ and $\theta$ are solutions of the equations $R=0$ and $\Theta=0$ respectively.
To find out which photons actually reach the observer one thus must find those pairs $(b,q^2)$ that satisfy Eq. (\ref{eqrth}). 

It is useful to introduce the new variable $\mu=\cos\theta$, so that the null geodesic equations (\ref{nullgeos}) become 
\begin{eqnarray}
\label{nullgeos2a}
&&\epsilon_r\int^{r}\frac{\rmd r}{\sqrt{R(r)}}=
\epsilon_\mu\int^{\mu}\frac{\rmd \mu}{\sqrt{\Theta_\mu(\mu)}}\ , \\
\label{nullgeos2b}
&&t=\epsilon_r\int^{r}\frac{r^2}{N^2\sqrt{R(r)}}\,\rmd r\ , \\
\label{nullgeos2c}
&&\phi=b\epsilon_\mu\int^{\mu}\frac{1}{1-\mu^2}\frac{\rmd \mu}{\sqrt{\Theta_\mu(\mu)}}\ ,
\end{eqnarray}
where
\beq\fl\qquad
\Theta_\mu=q^2-(q^2+b^2)\mu^2\equiv (q^2+b^2)(\bar\mu^2-\mu^2)\ , \qquad
\bar\mu^2=\frac{q^2}{q^2+b^2}\ .
\eeq
We will consider the case of an emitting source moving on the equatorial plane (i.e. $\theta_{\rm em}=\pi/2$) and a distant observer located far away from the black hole (i.e. $r_{\rm obs}\to\infty$) at azimuthal position $\phi_{\rm obs}=0$.
For a photon emitted by the spot we thus have $\mu_{\rm em}=0$. 
Furthermore, for a photon crossing the equatorial plane (which is the case we are interested in), we have $q^2 > 0$, so that $\bar\mu^2>0$.

Consider first Eq. (\ref{nullgeos2a}).
The integral over $\mu$ is straightforward  
\beq
\label{muint}
\int^\mu\frac{\rmd \mu}{\sqrt{\Theta_\mu}}=\frac{1}{\sqrt{q^2+b^2}}\arctan\left(\frac{\mu}{\sqrt{\bar\mu^2-\mu^2}}\right)\ .
\eeq

The integral over $r$ can be worked out with the inverse Jacobian elliptic integrals \cite{BF}.
Let us denote the four roots of $R(r)=0$ by $r_1$, $r_2$, $r_3$ and $r_4$.
There are two relevant cases to be considered.

\begin{enumerate}

\item[Case A:] $R(r) = 0$ has four real roots.

Let the roots be ordered so that $r_1\ge r_2\ge r_3\ge r_4$, with $r_3=0$ and $r_4\le0$.
Physically allowed regions for photons are given by $R\ge0$, i.e. $r\ge r_1$ (region I) and $r_3\le r\le r_2$ (region II) with $r>2M$.

In region I the integral over $r$ can be worked out by the following integration
\beq
\label{caseAint1}
\int_{r_1}^r\frac{\rmd r}{\sqrt{R(r)}}=\frac{2}{\sqrt{r_1(r_2-r_4)}}\,{\rm sn}^{-1}\left(\sin\varphi_{A_I}\vert k_A\right)\ ,
\eeq
where 
\beq
\sin\varphi_{A_I}=\sqrt{\frac{(r_2-r_4)(r-r_1)}{(r_1-r_4)(r-r_2)}}\ , \qquad
k_A=\sqrt{\frac{r_2(r_1-r_4)}{r_1(r_2-r_4)}}\ , 
\eeq
when $r_1\not=r_2$.
The case of two equal roots $r_1=r_2$ should be treated separately.

In region II the integral over $r$ can be worked out by the following integration
\beq
\label{caseAint2}
\int_r^{r_2}\frac{\rmd r}{\sqrt{R(r)}}=\frac{2}{\sqrt{r_1(r_2-r_4)}}\,{\rm sn}^{-1}\left(\sin\varphi_{A_{II}}\vert k_A\right)\ ,
\eeq
where
\beq
\sin\varphi_{A_{II}}=\sqrt{\frac{r_1(r_2-r)}{r_2(r_1-r)}}\ , \qquad
r_1\not=r_2\ .
\eeq

\item[Case B:] $R(r) = 0$ has two complex roots and two real roots.

Let us assume that $r_1$ and $r_2$ are complex, with $r_1={\bar r_2}$, whereas $r_3=0$ and $r_4$ is real such that $r_4\le0$.
The physically allowed region for photons is given by $r>2M$.

The integral over $r$ can be worked out with the following integration
\beq
\label{caseBint1}
\int_{2M}^r\frac{\rmd r}{\sqrt{R(r)}}=\int_{0}^r\frac{\rmd r}{\sqrt{R(r)}}-\int_0^{2M}\frac{\rmd r}{\sqrt{R(r)}}\ ,
\eeq
with
\begin{eqnarray}
\int_{0}^r\frac{\rmd r}{\sqrt{R(r)}}&=&\frac{1}{\sqrt{AB}}\,{\rm cn}^{-1}\left(\sin\varphi_{B}\vert k_B\right)\ , \nonumber\\
\int_0^{2M}\frac{\rmd r}{\sqrt{R(r)}}&=&\frac{1}{\sqrt{AB}}\,{\rm cn}^{-1}\left(\sin\varphi_{B}^H\vert k_B\right)\ ,
\end{eqnarray}
where 
\beq
\label{ABdefs}
A^2=u^2+v^2\ , \qquad
B^2=(r_4-u)^2+v^2\ , 
\eeq
with $u={\rm Re}(r_1)$ and $v={\rm Im}(r_1)$, and
\beq
\sin\varphi_{B}=\frac{(A-B)r-r_4A}{(A+B)r-r_4A}\ , \qquad
k_B=\sqrt{\frac{(A+B)^2-r_4^2}{4AB}}\ ,
\eeq
with $\sin\varphi_{B}^H=\sin\varphi_{B}(r=2M)$.

\end{enumerate}

\section{Images}

The apparent position of the image of the emitting source on the celestial sphere is represented by two impact parameters, $\alpha$ and $\beta$, measured on a plane centered about the observer location and perpendicular to the direction $\theta_{\rm obs}$. 
They are defined by \cite{cunningham}
\begin{eqnarray}
\label{impactparams}
\alpha&=&\lim_{r_{\rm obs}\to\infty}-r_{\rm obs}\frac{K^{\hat \phi}}{K^{\hat t}}
=-\frac{b}{\sin\theta_{\rm obs}}\ , \nonumber\\
\beta&=&\lim_{r_{\rm obs}\to\infty}r_{\rm obs}\frac{K^{\hat \theta}}{K^{\hat t}}
=\epsilon_{\theta_{\rm obs}}\sqrt{q^2-b^2\cot^2\theta_{\rm obs}}\ ,
\end{eqnarray}
where $K^{\hat \alpha}$ are the frame components of $K$ with respect to ZAMOs (its coordinate components are instead listed in Eq. (\ref{Kcompts})).
Equivalent expressions can be obtained by decomposing the photon 4-velocity as follows
\beq
K=E(n)[e_{\hat t}+{\hat \nu}^{\hat a}e_{\hat a}]\ , \quad {\hat \nu}\cdot{\hat \nu}=1\ ,
\eeq
so that 
\begin{eqnarray}
\alpha&=&\lim_{r_{\rm obs}\to\infty}-r_{\rm obs}{\hat \nu}^{\hat \phi}=\lim_{r_{\rm obs}\to\infty}[\vec r\times\hat \nu]^{\hat \theta}
\ , \nonumber\\
\beta&=&\lim_{r_{\rm obs}\to\infty}r_{\rm obs}{\hat \nu}^{\hat \theta}=\lim_{r_{\rm obs}\to\infty}[\vec r\times\hat \nu]^{\hat \phi}\ .
\end{eqnarray}

The line of sight to the black hole center marks the origin of the coordinates, where $\alpha=0=\beta$.
Now imagine a source of illumination behind the black hole whose angular size 
is large compared with the angular size of the black hole. As seen by the distant 
observer the black hole will appear as a black region in the middle of the larger 
bright source. No photons with impact parameters in a certain range about $\alpha=0=\beta$ 
will reach the observer. 
The rim of the black hole corresponds to photon 
trajectories which are marginally trapped by the black hole; they spiral around 
many times before they reach the observer. 

The image of the trajectory is thus obtained by determining all pairs $(b,q^2)$ satisfying Eq. (\ref{eqrth}) (or equivalently Eq. (\ref{nullgeos2a})), then substituting back into Eq. (\ref{impactparams}) in order to get the corresponding coordinates on the observer's plane.
Alternatively, one can solve Eq. (\ref{impactparams}) for $b$ and $q^2$, i.e.
\beq
\fl\quad
b=-\alpha\sin\theta_{\rm obs}\ , \qquad
q^2=\beta^2+\alpha^2\cos^2\theta_{\rm obs} \quad \to \quad
b^2+q^2=\alpha^2+\beta^2\ ,
\eeq
then substituting back into Eq. (\ref{nullgeos2a}) and solving for all allowed pairs of impact parameters $(\alpha,\beta)$.

The images of the source obtained in this way can be classified according to the number of times the photon trajectory crosses the equatorial plane between the emitting source and the observer.
The trajectory of the \lq\lq direct'' image does not cross the equatorial plane; that of a \lq\lq first-order'' image crosses it once; and so on.

We are interested in constructing direct images only and refer to \ref{image} for details.

\section{Light curves}

The observed differential flux is given by \cite{MTW}
\beq
\rmd F_{\rm obs}=I_{\rm obs}\rmd\Omega\ ,
\eeq
where $\rmd\Omega$ is the solid angle subtended by the light source on the observer sky and $I_{\rm obs}$ the intensity of the source integrated over its effective frequency range, i.e. 
\beq
I_{\rm obs}=g^4I_{\rm em}\ .
\eeq
The intensity $I_{\rm em}$ measured at the rest frame of the spot can be normalized as $I_{\rm em}=1$.
Furthermore, the solid angle can be expressed in terms of the observer's plane coordinates $\{\alpha, \beta\}$ as  
\beq
\rmd \Omega=\frac{\rmd\alpha\rmd\beta}{r_{\rm obs}^2}\ .
\eeq
Introducing then polar coordinates on the observer's plane and switching integration over $r_{\rm em}$ and $\phi_{\rm em}$ (see \ref{flux} for details), the observed differential flux is expressed as 
\beq\fl\qquad
\label{fluxfin}
\rmd F_{\rm obs}=\frac{g^4}{r_{\rm obs}^2}
\frac{\cos\theta_{\rm obs}}{1-\cos^2\phi_{\rm em}\sin^2\theta_{\rm obs}}
\left|b\frac{\partial b}{\partial r_{\rm em}}+q\frac{\partial q}{\partial r_{\rm em}}\right|
\rmd r_{\rm em}\rmd \phi_{\rm em}\ .
\eeq

Finally, the light curve of the emitting source is  constructed by introducing the time dependence of the radiation received by the distant observer, including the time delay effects.
Therefore, we also need to evaluate the coordinate time interval spent by each photon to reach the observer (see \ref{time}).

\section{Results}

\begin{figure} 
\typeout{*** EPS figure 1}
\begin{center}
\includegraphics[scale=0.5]{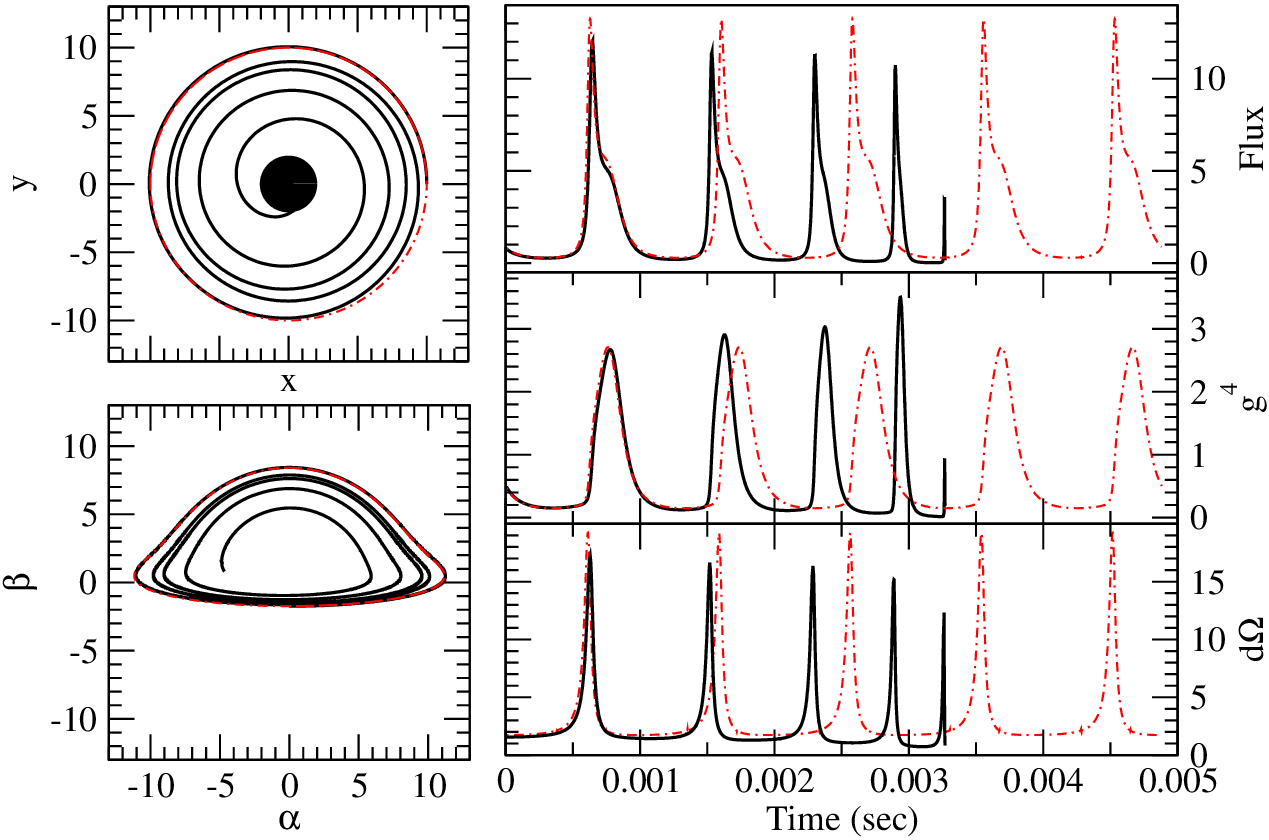}
\end{center}
\caption{
The apparent position (direct image only), light curve, redshift factor and solid angle of the emitting spot are shown   for the orbit depicted in the upper left panel in the case of a radially outgoing radiation field.
The orbital parameters and initial conditions are $A/M=0.01$, $r_{\rm em}(0)=10M$, $\phi_{\rm em}(0)=0$, $\nu_{\rm em}(0)=\nu_K\approx0.35$, $\alpha_{\rm em}(0)=0$. 
The distant observer is located at the polar angle $\theta_{\rm obs}=80^\circ$. 
$x=(r_{\rm em}/M)\cos\phi_{\rm em}$ and $y=(r_{\rm em}/M)\sin\phi_{\rm em}$ are Cartesian-like coordinates expressed in units of $M$.
The black circle represents the Schwarzschild horizon $r=2M$.
The critical radius approaches the horizon in this case.
The flux is given in arbitrary units as a function of the coordinate time given in seconds, corresponding to the choice of $M=1.0M_\odot$.
For comparison purposes, the corresponding curves for
an emitting spot in circular geodesic orbit at $r_{\rm em}=10M$
on the equatorial plane 
of a Schwarzschild spacetime are shown (dashed curves).
}
\label{fig:1}
\end{figure}

\begin{figure} 
\typeout{*** EPS figure 2}
\begin{center}
\includegraphics[scale=0.5]{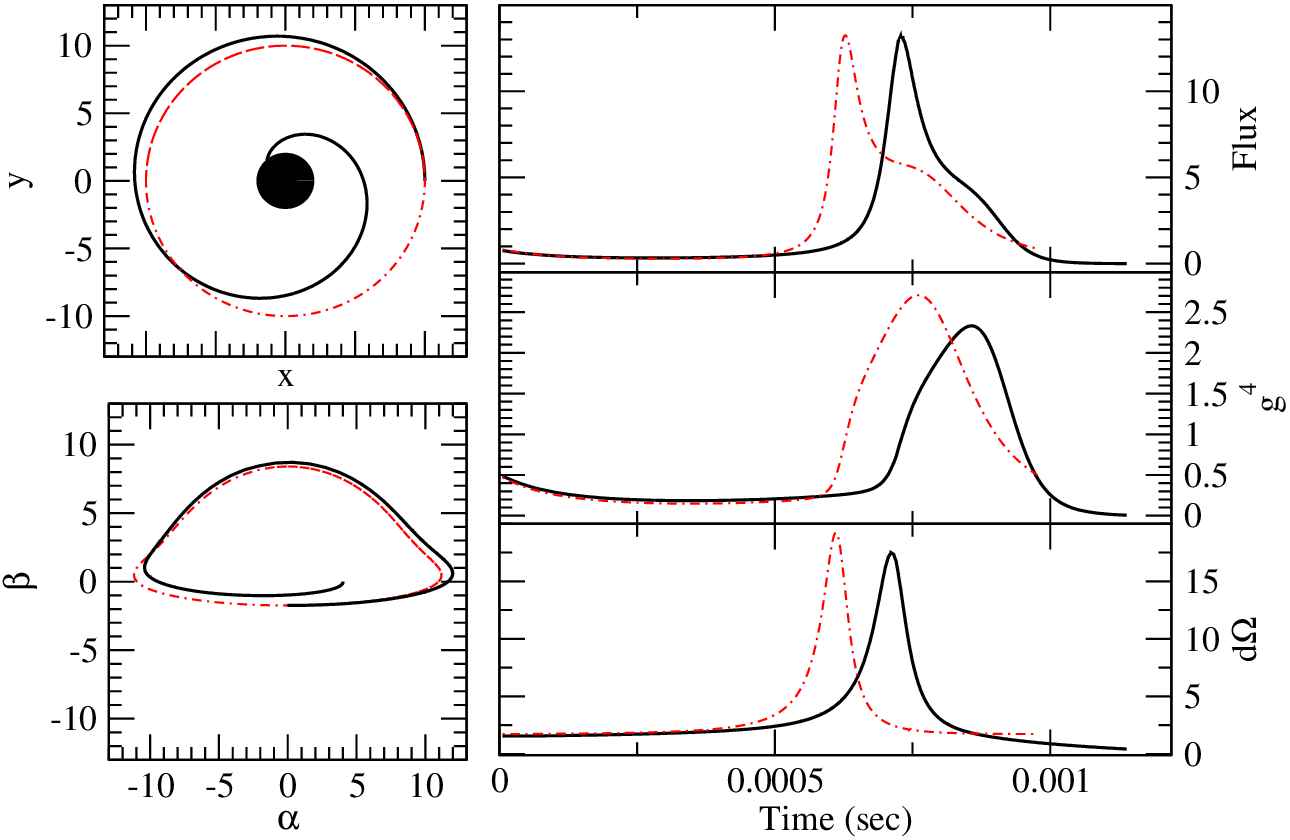}
\end{center}
\caption{
The same as in Fig. \ref{fig:1} but with $A/M=0.1$.
The critical radius approaches the horizon also in this case ($r_{\rm(crit)}\approx 2.02M$).
} 
\label{fig:2}
\end{figure}

\begin{figure} 
\typeout{*** EPS figure 3}
\begin{center}
\includegraphics[scale=0.5]{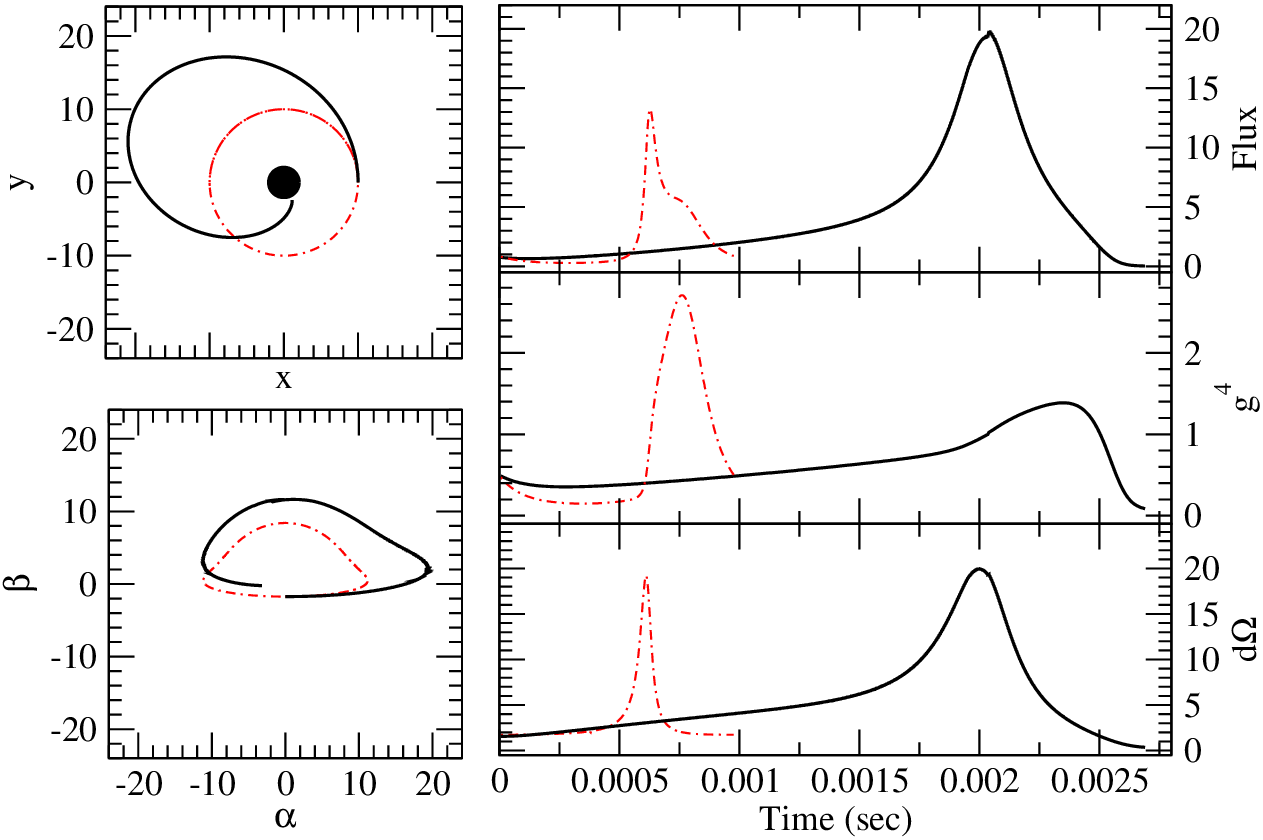}
\end{center}
\caption{
The same as in Fig. \ref{fig:1} but with $A/M=0.5$.
The critical radius is $r_{\rm(crit)}\approx2.67M$.
}
\label{fig:3}
\end{figure}

\begin{figure} 
\typeout{*** EPS figure 4}
\begin{center}
\includegraphics[scale=0.5]{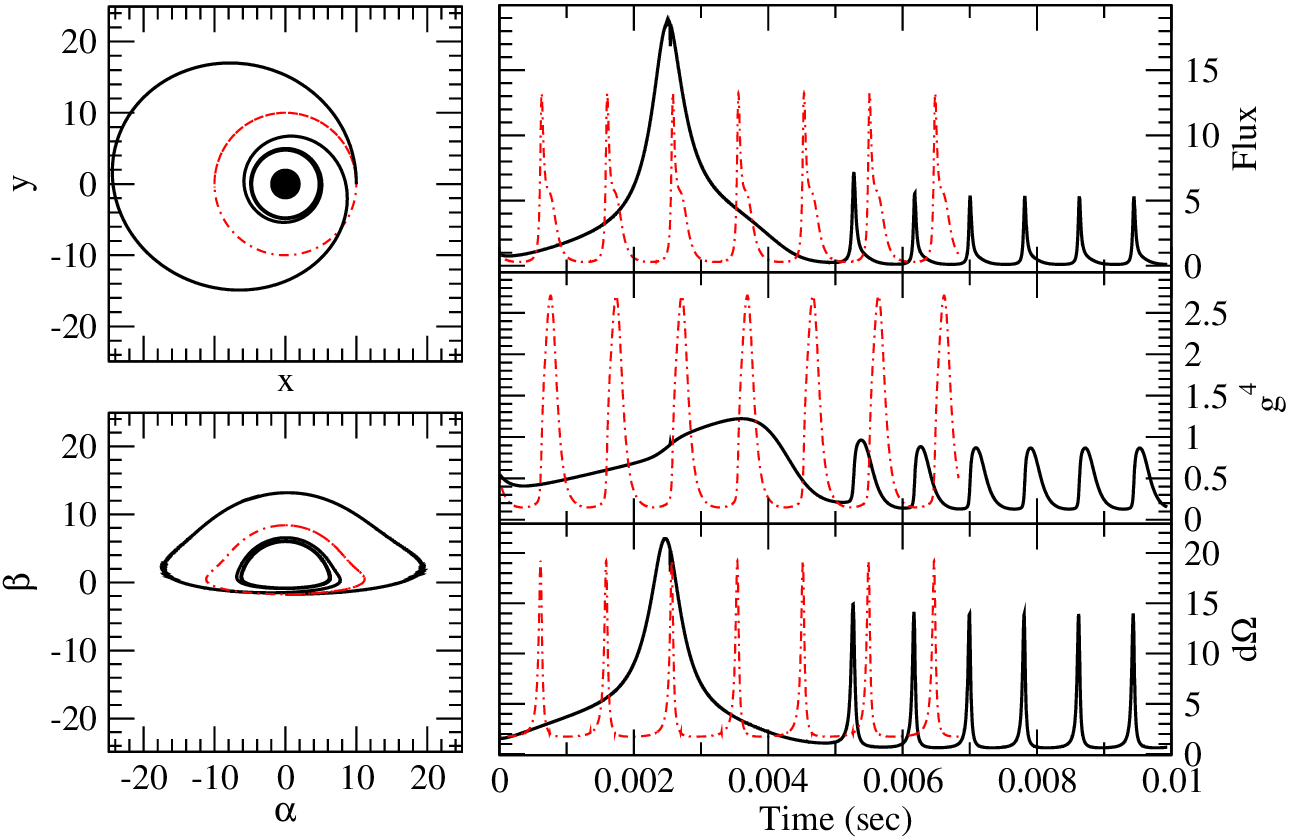}
\end{center}
\caption{
The same as in Fig. \ref{fig:1} but with luminosity parameter $A/M=0.7$ and angular momentum $b_{\rm(rad)}=1.5M$ of the radiation field and initial velocity $\nu_{\rm em}(0)=0.28$.
The critical radius is $r_{\rm(crit)}\approx4.76M$ and the critical velocity is $\nu_{\rm(crit)}\approx0.24$.
Note that the Keplerian velocity at that radius would be $0.60$.
}
\label{fig:4}
\end{figure}

\begin{figure} 
\typeout{*** EPS figure 5}
\begin{center}
\includegraphics[scale=0.5]{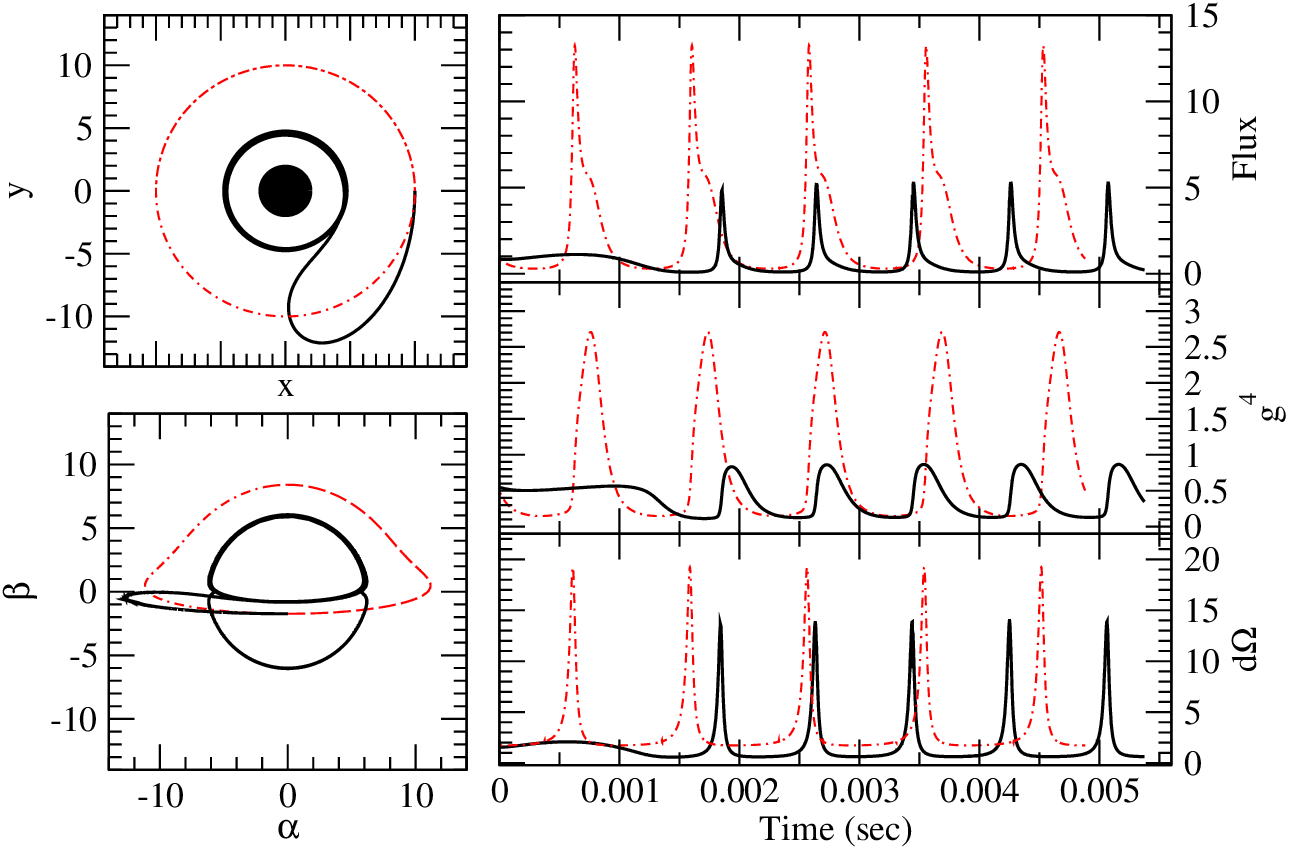}
\end{center}
\caption{
The same as in Fig. \ref{fig:4} but with an initial negative value of the azimuthal velocity, i.e. $\alpha_{\rm em}(0)=\pi$.
The values of the critical radius as well as critical velocity do not change.
}
\label{fig:5}
\end{figure}

Figures \ref{fig:1}--\ref{fig:5} show the apparent position of the direct image, light curve, redshift factor and solid angle for an emitting spot under the effect of both the gravitational and radiation fields. Different figures correspond to different properties of the radiation field.
The distant observer is located at the polar angle $\theta_{\rm obs}=80^\circ$ in all cases.
The case of an emitting spot in circular geodesic motion on the equatorial plane of a Schwarzschild spacetime at the same initial $r_{\rm em}$ is also shown for comparison. 

In Figs. \ref{fig:1}--\ref{fig:3} the initial radius and azimuthal angle of the emitting spot are $r_{\rm em}(0)=10M$, $\phi_{\rm em}(0)=0$; the initial velocity is that of a Keplerian circular orbit at that radius.
The radiation field is radially outgoing, with different values of the luminosity parameter $A/M$.   
For small values of $A/M$ the emitter spirals towards the critical radius located close to the horizon, by undergoing several azimuthal cycles around the black hole (see Fig. \ref{fig:1}).
The trajectory is nearly circular after the first revolution, so that the light curve is very close to the circular Keplerian one.
The spiraling then becomes faster, and the light curve peaks occur faster and faster.
 
Increasing the luminosity parameter while maintaining the initial conditions fixed causes the emitting spot to drift initially to larger radii and then quickly spiral inwards down to the critical radius, which as expected is larger in this case (see Figs. \ref{fig:2} and \ref{fig:3} and Eq. (\ref{rcrit_zeroang})). 

Figures \ref{fig:4} and \ref{fig:5} show the effect of a nonzero angular momentum of the radiation field.
In the case of Fig. \ref{fig:4} the emitting spot drifts away from the black hole before going back inwards, ending in the circular equilibrium orbit in a few revolutions. 
This is also evident from the comparison of the corresponding light curve with the Keplerian one.
In fact, the emitter spends fairly long time far from the source before completing the first revolution, whereas in the same time interval the Keplerian emitter orbits the source  several times.
Then the orbit becomes soon circular,  at the  smaller radius given by Eq. (\ref{forcebalance2s}).
Fig. \ref{fig:5} refers to an emitter with the same initial conditions as in Fig. \ref{fig:4}, except for the negative sign of the azimuthal velocity.
In this case, the emitter initially moves 
clockwise around the black hole and is quickly dragged by the radiation field in the opposite direction, soon reaching the same equilibrium orbit as in Fig. \ref{fig:4}.


\section{Concluding remarks} 

We have calculated the signal produced by an emitting point-like source moving in the equatorial plane of a Schwarzschild spacetime under the influence of a  radiation field. 
The latter consists of photons having the same specific angular momentum and traveling along geodesics.
The interaction with the photon field leads to a friction-like drag force responsible for the so called Poynting-Robertson effect. Previous studies have shown that, in the case of photons with zero angular momentum, i.e., propagating radially with respect to the ZAMOs, there exists  an equilibrium radius representing the balance of the outward radiation force with the inward gravitational force, where the emitter can remain at rest.
The location of such a critical radius depends on the luminosity parameter.
If the outward photon flux possesses a non-zero angular momentum, emitting spots that do not escape end up in circular orbits.
In the present paper we have derived the flux, redshift factor and solid angle as a function of the (coordinate) time, as well as the time-integrated image of the spot in the observer's sky.
The results are clearly different from those for an emitting spot in circular geodesic motion, as shown by numerical examples where the effect of the interaction with the radiation field has been investigated by varying both the luminosity parameter and the photon angular momentum.

The treatment and results presented here hold a potential for astrophysical applications. 
Matter accretion towards white dwarfs, neutron stars or black holes that emit radiation at a sizeable fraction of their Eddington luminosity (i.e., $L/L_{Edd}\gtrsim1\%$) will be influenced by general relativistic Poynting-Robertson-type effects. Departures from the unperturbed motion (i.e., in the absence of the radiation field) are substantial and may lead to observable phenomena. The range of astrophysical applications is vast; examples are 
quasi periodic oscillations (QPOs) that are observed in accreting neutron stars and black holes \cite{vanderklis95, mchardy}, thermonuclear flashes that occur on the surface of accreting neutron stars (the so called type-I bursts \cite{strohmayer}) and the very broad Fe-K$\alpha$ line profiles produced in
the innermost regions of accretion disks around collapsed objects \cite{citare}. However, the treatment we have developed here is idealized in several aspects and the impact of some approximations should be carefully assessed before detailed predictions for astrophysical systems are worked out. Treating the photon field as if all photons had the same angular momentum presents clear advantages for the analytical calculations presented in Refs. \cite{bijanste,pr2}, but would require some caution in an astrophysical context. For instance, the flux emitted from the surface of an accreting neutron star comprises photons emitted in virtually all directions (and thus possessing a range of different angular momenta). Moreover, in accreting black holes the motion of matter in the vicinity of the innermost stable circular orbit (ISCO) will be mainly affected by radiation coming from the outer disk regions, in turn involving a radiation field emitted in a range of different photon directions and emission radii. Finally, in an astrophysical environment one must also consider the impact of two key assumptions, which are intrinsic to any Poynting-Robertson-type theory, namely that matter is directly exposed to the radiation field (meaning that the optical depth to the source must be $<1$) and that the radiation re-emitted or scattered by matter propagates unimpeded without undergoing other interactions. 

Despite these limitations the analysis presented here captures some essential features of the motion of matter in the strong field regime under the effects of an intense source of radiation. Future work will be devoted to generalizing the present treatment and addressing specific astrophysical situations in which the general relativistic version of the Poynting-Robertson effect is relevant.

\appendix

\section*{Appendix}

We list below for completeness the details on the construction of the direct image, the derivation of the observed energy flux and the calculation of the coordinate time interval between emitter and observer.
This is a well known topic addressed by many authors in the literature (see, e.g., Refs. \cite{rauch,bao,lietal,agol,gyoto}).
However, different and sophisticated techniques are used simply to show light curves as well as images, without entering the underlying analytical framework or referring to previous related works.

\section{Constructing the direct image}
\label{image}

The direct image results from photons which never cross the equatorial plane.
As the photon reaches the observer, on the photon orbit we have $\rmd\theta/\rmd r > 0$ (i.e. $\rmd\mu/\rmd r < 0$) if $\beta > 0$, and $\rmd\theta/\rmd r < 0$ (i.e. $\rmd\mu/\rmd r > 0$) if $\beta < 0$. 
Therefore, when $\beta > 0$ the photon must encounter a turning point at $\mu=\bar\mu$: $\mu$ starts from 0, goes up to $\bar\mu$, then goes down to $\mu_{\rm obs}$ (which is $\le\bar\mu$).
When $\beta < 0$, the photon do not encounter a turning point at $\mu=\bar\mu$: $\mu$ starts from 0 and monotonically increases to $\mu_{\rm obs}$.

The total integration over $\mu$ along the path of the photon from the emitting source to the observer is thus given by
\beq
\label{Imudir}
I_\mu=\left\{
\begin{array}{lr}
\left[\int_0^{\bar\mu}+\int_{\mu_{\rm obs}}^{\bar\mu}\right]\frac{\rmd \mu}{\sqrt{\Theta_\mu(\mu)}}&(\beta > 0)\\ 
&\\
\int_0^{\mu_{\rm obs}}\frac{\rmd \mu}{\sqrt{\Theta_\mu(\mu)}}=\left[\int_0^{\bar\mu}-\int_{\mu_{\rm obs}}^{\bar\mu}\right]\frac{\rmd \mu}{\sqrt{\Theta_\mu(\mu)}}&(\beta < 0)\ .
\end{array}
\right.
\eeq
By using Eq. (\ref{muint}) we get
\beq
\label{muintdirect}
I_\mu=\left\{
\begin{array}{lr}
\frac{\pi}{\sqrt{\alpha^2+\beta^2}}-I_{\mu_{\rm obs}}&(\beta > 0)\\
&\\
I_{\mu_{\rm obs}}&(\beta < 0)\ ,
\end{array}
\right.
\eeq
where
\beq
I_{\mu_{\rm obs}}=\frac{1}{\sqrt{\alpha^2+\beta^2}}\arctan\left(\frac{\sqrt{\alpha^2+\beta^2}}{|\beta|}\cot\theta_{\rm obs}\right)\ .
\eeq

Now let us consider the integration over $r$.
Since the observer is at infinity, the photon reaching him/her must have been moving in the allowed region defined by $r\ge r_1$ when $R(r)=0$ has four real roots (case A), or the allowed region defined by $r>2M$ when $R(r)=0$ has two complex roots and two real roots (case B). There are then two possibilities for the photon during its trip: it has encountered a turning point at $r=r_1$, or it has not encountered any turning point in $r$. Define
\beq
\label{defsIr}
I_\infty\equiv\int_{r_t}^\infty\frac{\rmd r}{\sqrt{R(r)}}\ , \qquad
I_{r_{\rm em}}\equiv\int_{r_t}^{r_{\rm em}}\frac{\rmd r}{\sqrt{R(r)}}\ .
\eeq
Obviously, according to Eq. (\ref{nullgeos2a}), a necessary and sufficient condition for the occurrence of a turning point in $r$ on the path of the photon is that $I_\infty<I_\mu$.
Therefore, the total integration over $r$ along the path of the photon from the emitting source to the observer is
\beq
\label{Irdir}
I_r=\left\{
\begin{array}{lr}
I_\infty+I_{r_{\rm em}}&(I_\infty<I_\mu)\\
&\\
\int_{r_{\rm em}}^\infty\frac{\rmd r}{\sqrt{R(r)}}=I_\infty-I_{r_{\rm em}}&(I_\infty\ge I_\mu)\ .
\end{array}
\right.
\eeq
By definition, $I_\infty$, $I_{r_{\rm em}}$, and $I_r$ are all positive. According to Eq. (\ref{nullgeos2a}), we must have $I_r=I_\mu$ for the orbit of a photon.
The relevant cases to be considered are the following.

\begin{enumerate}

\item[Case A:] $R(r) = 0$ has four real roots.

When $r_1\not=r_2$, by using Eq. (\ref{caseAint1}) to evaluate integrals in Eq. (\ref{defsIr}) with $r_t=r_1$ we get 
\begin{eqnarray}
I_\infty&=&\frac{2}{\sqrt{r_1(r_2-r_4)}}\,{\rm sn}^{-1}\left(\sin\varphi_{A_I}^\infty\vert k_A\right)\ , \nonumber\\
I_{r_{\rm em}}&=&\frac{2}{\sqrt{r_1(r_2-r_4)}}\,{\rm sn}^{-1}\left(\sin\varphi_{A_I}^{\rm em}\vert k_A\right)\ ,
\end{eqnarray}
where $\sin\varphi_{A_I}^\infty=\sin\varphi_{A_I}(r\to\infty)$ and $\sin\varphi_{A_I}^{\rm em}=\sin\varphi_{A_I}(r=r_{\rm em})$.
Substitute these expressions into Eq. (\ref{Irdir}), then let $I_r=I_\mu$ and finally solve for $r_{\rm em}$:
\beq
\label{remcaseA}
r_{\rm em}=\frac{r_1(r_2-r_4)-r_2(r_1-r_4){\rm sn}^2(\sin\xi_A\vert k_A)}{(r_2-r_4)-(r_1-r_4){\rm sn}^2(\sin\xi_A\vert k_A)}\ ,
\eeq
where
\beq
\sin\xi_A=\frac12(I_\mu-I_\infty)\sqrt{r_1(r_2-r_4)}\ .
\eeq

Since ${\rm sn}^2(-\sin\xi_A\vert k_A)={\rm sn}^2(\sin\xi_A\vert k_A)$, the solution given by Eq. (\ref{remcaseA}) applies whether $I_\mu-I_\infty$ is positive or negative, i.e. no matter whether there is a turning point in $r$ or not along the path of the photon.

\item[Case B:] $R(r) = 0$ has two complex roots and two real roots.

No turning points occur in this case.
Therefore, we have
\beq
I_r=\int_{r_{\rm em}}^\infty\frac{\rmd r}{\sqrt{R(r)}}=I_\infty-I_{r_{\rm em}}\ ,
\eeq
with 
\beq\fl\quad
I_\infty=\frac{1}{\sqrt{AB}}\,{\rm cn}^{-1}\left(\sin\varphi_{B}^\infty\vert k_B\right)\ , \qquad
I_{r_{\rm em}}=\frac{1}{\sqrt{AB}}\,{\rm cn}^{-1}\left(\sin\varphi_{B}^{\rm em}\vert k_B\right)\ ,
\eeq
by using Eq. (\ref{caseBint1}) to evaluate integrals in Eq. (\ref{defsIr}), also recalling Eq. (\ref{ABdefs}).
Solving then the equation $I_r=I_\mu$ for $r_{\rm em}$ gives
\beq
\label{remcaseB}
r_{\rm em}=\frac{r_4A[1-{\rm cn}(\sin\xi_B\vert k_B)]}{(A-B)-(A+B){\rm cn}(\sin\xi_B\vert k_B)}\ ,
\eeq
where $\sin\varphi_{B}^\infty=\sin\varphi_{B}(r\to\infty)$ and $\sin\varphi_{B}^{\rm em}=\sin\varphi_{B}(r=r_{\rm em})$ and
\beq
\sin\xi_B=(I_\mu-I_\infty)\sqrt{AB}\ .
\eeq

\end{enumerate}

\section{Evaluating the observed energy flux}
\label{flux}

The solid angle once expressed in terms of the observer's plane coordinates $\{\alpha, \beta\}$ is given by  
\beq
\label{solid}
\rmd \Omega=\frac{\rmd\alpha\rmd\beta}{r_{\rm obs}^2}\ .
\eeq
Introduce polar coordinates on the observer's plane
\beq
\alpha=\rho\cos\psi\ , \qquad
\beta=\rho\sin\psi\ .
\eeq
The integration over the observer's plane coordinates can then be switched over $r_{\rm em}$ and $\phi_{\rm em}$ by
\beq
\rmd\alpha\rmd\beta=\rho\,\rmd\rho\rmd\psi
=\sqrt{\alpha^2+\beta^2}\left|\frac{\partial(\rho,\psi)}{\partial(r_{\rm em},\phi_{\rm em})}\right|\rmd r_{\rm em}\rmd \phi_{\rm em}\ ,
\eeq
where
\beq
\left|\frac{\partial(\rho,\psi)}{\partial(r_{\rm em},\phi_{\rm em})}\right|
=\left|\frac{\partial \rho}{\partial r_{\rm em}}\frac{\partial \psi}{\partial \phi_{\rm em}}-\frac{\partial \rho}{\partial \phi_{\rm em}}\frac{\partial \psi}{\partial r_{\rm em}}\right|\ 
\eeq
is the Jacobian of the transformation $(\rho,\psi)\to(r_{\rm em},\phi_{\rm em})$.
Since 
\beq
\rho=\sqrt{\alpha^2+\beta^2}=\sqrt{b^2+q^2}\ 
\eeq
does not depend on $\phi_{\rm em}$, we have to evaluate only
\beq
\frac{\partial \rho}{\partial r_{\rm em}}=\frac{1}{\rho}\left(b\frac{\partial b}{\partial r_{\rm em}}+q\frac{\partial q}{\partial r_{\rm em}}\right)\ ,
\eeq
where the derivatives of $b$ and $q$ with respect to $r_{\rm em}$ are obtained simply by inverting the derivatives $\partial r_{\rm em}/\partial q$ and $\partial r_{\rm em}/\partial b$ which can be evaluated from Eqs. (\ref{remcaseA}) and (\ref{remcaseB}).

$\partial\psi/\partial\phi_{\rm em}$ instead can be evaluated from Eq. (\ref{nullgeos2c}) governing the azimuthal motion, i.e.
\beq
\phi=b\epsilon_\mu\int^{\mu}\frac{1}{1-\mu^2}\frac{\rmd \mu}{\sqrt{\Theta_\mu(\mu)}}\ ,
\eeq 
taking into account that $\phi_{\rm obs}=0$ and $\tan\psi=\beta/\alpha$.
The integration is straightforward
\beq
\phi=-\epsilon_\mu \arctan\left(b\frac{\mu}{\sqrt{\Theta_\mu(\mu)}}\right)\ ,
\eeq
so that 
\beq
\phi_{\rm em}=\left\{
\begin{array}{lr}
\pi+\arctan\left(\displaystyle\frac{\alpha}{\beta}\cos\theta_{\rm obs}\right)&(\beta > 0)\\
-\arctan\left(\displaystyle\frac{\alpha}{|\beta|}\cos\theta_{\rm obs}\right)&(\beta < 0)\ ,
\end{array}
\right.
\eeq
whence
\beq
\tan\phi_{\rm em}=\frac{\alpha}{\beta}\cos\theta_{\rm obs}=\cot\psi\cos\theta_{\rm obs}\ .
\eeq
Therefore
\beq\fl\qquad
\sin\psi=\frac{\cos\phi_{\rm em}\cos\theta_{\rm obs}}{\sqrt{1-\cos^2\phi_{\rm em}\sin^2\theta_{\rm obs}}}\ , \quad
\cos\psi=\frac{\sin\phi_{\rm em}}{\sqrt{1-\cos^2\phi_{\rm em}\sin^2\theta_{\rm obs}}}\ , 
\eeq
implying that 
\beq
\frac{\partial \psi}{\partial \phi_{\rm em}}=\frac{\cos\theta_{\rm obs}}{1-\cos^2\phi_{\rm em}\sin^2\theta_{\rm obs}}\ .
\eeq
Finally, the solid angle (\ref{solid}) turns out to be given by
\beq\fl\qquad
\rmd \Omega=\frac{1}{r_{\rm obs}^2}
\frac{\cos\theta_{\rm obs}}{1-\cos^2\phi_{\rm em}\sin^2\theta_{\rm obs}}
\left|b\frac{\partial b}{\partial r_{\rm em}}+q\frac{\partial q}{\partial r_{\rm em}}\right|
\rmd r_{\rm em}\rmd \phi_{\rm em}\ ,
\eeq
leading to the expression (\ref{fluxfin}) for the observed differential flux.

\section{Evaluating the coordinate time integral}
\label{time}

The light travel time between emitter and observer is given by Eq. (\ref{nullgeos2b}), i.e. 
\beq
t_{\rm obs}-t_{\rm em}=\epsilon_r\int^{r}\frac{r^2}{N^2\sqrt{R(r)}}\,\rmd r\bigg\vert_{r_{\rm em}}^{r_{\rm obs}}\ ,
\eeq
where the integration has to be done properly.
The integral can be conveniently decomposed as follows
\begin{eqnarray}
\label{tint}
\fl\qquad
\int^{r}\frac{r^2}{N^2\sqrt{R(r)}}\,\rmd r&=&\int^{r}\frac{r^3}{(r-2M)\sqrt{R(r)}}\,\rmd r\nonumber\\
\fl\qquad
&=&8M^3\int^{r}\frac{\rmd r}{(r-2M)\sqrt{R(r)}}
+\int^{r}\frac{r^2}{\sqrt{R(r)}}\,\rmd r\nonumber\\
\fl\qquad
&&+2M\int^{r}\frac{r}{\sqrt{R(r)}}\,\rmd r
+4M^2\int^{r}\frac{\rmd r}{\sqrt{R(r)}}\ . 
\end{eqnarray}
Each term can be evaluated in terms of elliptic functions, e.g., by using the table of integrals in Ref. \cite{BF}. 

Consider first the Case A, where $R(r) = 0$ has four real roots.
Let the roots be ordered so that $r_1> r_2> r_3> r_4$, with $r_4<0$.
Physically allowed regions for photons are given by $R>0$, i.e. $r>r_1$ (region I) and $r_2> r> r_3$ (region II).
In region I the integrals entering Eq. (\ref{tint}) have to be worked out by the formulas nn 258.00, 258.11, 258.39 on pp 128-132 of Ref. \cite{BF}.
In region II instead we refer to nn 255.00, 255.17, 255.38 on pp 116-120. 

In the Case B the equation $R(r) = 0$ has two complex roots and two real roots.
Let us assume that $r_1$ and $r_2$ are complex, $r_3$ and $r_4$ are real and $r_3 \ge r_4$. Then, we must have $r_1={\bar r_2}$, whereas $r_3\ge0$ and $r_4\le0$.
The physically allowed region for photons is given by $r> r_3$.
The integrals entering Eq. (\ref{tint}) have to be worked out by the formulas nn 260.00, 260.03, 260.04 on pp 135-136  of Ref. \cite{BF}.

\section*{Acknowledgement}

This work was partially supported through ICRANet and PRIN INAF 2008 contracts. LS and AG acknowledge the International Space Science Institute (ISSI) in Bern for the hospitality during part of this work was carried out. All the authors are indebted to Prof. R.T. Jantzen for stimulating discussions about the Poynting-Robertson effect in general relativity.

\end{document}